\documentstyle[12pt,epsfig]{article}
\renewcommand{\arraystretch}{1.5}
\def \gevcc {\mbox{GeV/$c^2$}}
\def \gevc  {\mbox{GeV/$c$}}
\def \mevcc {\mbox{MeV/$c^2$}}
\def \mevc  {\mbox{MeV/$c$}}

\def \bbbar {\mbox{$b{\bar b}$}}
\def \ccbar {\mbox{$c{\bar c}$}}
\def \micron {\mbox{$\mu$m}}

\def \dords {\mbox{$D^{(\star)}$}}
\def \aleph {\mbox{\sc Aleph}}

\def \fsp {\mbox{$f^\star_-$}}
\def \fso {\mbox{$f^\star_0$}}
\def \fsbg {\mbox{$f^\star_{\rm bkg}$}}
\def \fop {\mbox{$f^0_-$}}
\def \foo {\mbox{$f^0_0$}}
\def \fobg {\mbox{$f^0_{\rm bkg}$}}
\def \rphi {\mbox{$r\phi$}}

\def \d0 {\mbox{$D^0$}}
\def \dz {\mbox{$D^0$}}

\def \dkpi {\mbox{$D^0\rightarrow K^-\pi^+$}}
\def \pz  {\mbox{$\pi^0$}}
\def \dkppz {\mbox{$D^0\rightarrow K^-\pi^+\pi^0$}}
\def \dkpps {\mbox{$D^0\rightarrow K^0_{\rm S}\pi^+\pi^-$}}
\def \dkppp {\mbox{$D^0\rightarrow K^-\pi^+\pi^-\pi^+$}}

\def \dsdpi {\mbox{$\ds\rightarrow \d0 \pi^+$}}

\def\be {\begin{equation}}
\def\ee {\end{equation}}

\def\delm {\mbox{$m_{D^0\pi}-m_{D^0}$}}

\def\bbbar {\mbox{$b{\bar b}$}}

\def\dedx {\mbox{$dE/dx$}}

\def\dsl {\mbox{$D^{\star +} \ell^-$}}

\def\dl {\mbox{$D^{0} \ell^-$}}
\def\ds {\mbox{$D^{\star +}$}}

\def\r0 {\mbox{$R^0$}}

\def\b0 {\mbox{${\bar B}^0$}}

\def\bp {\mbox{$B^-$}}

\def\t0 {\mbox{$\tau_{\bar B^0}$}}
\def\tp {\mbox{$\tau_{B^-}$}}
\def\tBs {\mbox{$\tau_{B_s}$}}
\def\tLamb {\mbox{$\tau_{\Lambda_b}$}}

\def\chisq {\mbox{$\chi^2$}} 
\def\dpil  {\mbox{$\dords \pi \ell^- \nu$}}

\newcommand{\A}{{\sc Aleph}}

\newcommand{\ra}{\rightarrow}
\newcommand{\rarr}{\rightarrow}
\begin{document}
 
\thispagestyle{empty}
 
\centerline{{\Large EUROPEAN ORGANIZATION FOR NUCLEAR RESEARCH}} 
\vspace{9mm}   % da mettere al posto del 6mm quando tolgo il preliminary
%\vspace{4mm}
\noindent
\begin{flushright}
\vspace{2mm}
{ CERN-EP/2000-106}\\
July 26, 2000 \\
%\vskip 0.3 cm
\end{flushright}
\vskip 4 mm
\vspace{9mm}     %da mettere quando levo il preliminary (come sopra)
\begin{center}
\boldmath{
{\LARGE \bf Measurement of the \b0 \ and   \\[1ex]
\bp\ Meson Lifetimes}}
\vspace{10mm} \\
{\Large
{The ALEPH Collaboration}\\
\vspace{3mm}}
\vspace{6mm}
%{\LARGE PRELIMINARY}
%\vspace{8mm} \\
{\Large \bf Abstract}\end{center}
\begin{quote}
The lifetimes of the \b0 \ and \bp\ mesons are measured
using a sample of about four million hadronic Z decays collected from 1991 to 1995 
with the \aleph\ detector at LEP. The data sample has been recently 
reprocessed, achieving a substantial improvement in the tracking performance.
Semileptonic decays of \b0 \ and \bp\ mesons are
partially reconstructed by identifying events containing a lepton
with an associated \ds\ or \d0 \ meson. The proper time of the $B$ meson
is estimated from the measured decay length and the momentum 
of the $D$-lepton system. 
A fit to the proper time of 1880 $D^{\star+}\ell^{-}$ and 2856 $D^0\ell ^-$ candidates
yields the following results:
\vspace{4mm}
\begin{eqnarray*}
\t0 & =  & 1.518 \pm 0.053\pm 0.034 \mathrm{\ ps,} \\
\tp & =  & 1.648 \pm 0.049\pm 0.035 \mathrm{\ ps,} \\
\tp / \t0 & =  & 1.085 \pm 0.059 \pm 0.018.
\end{eqnarray*}
\end{quote}
\setlength{\textheight}{24.0cm}
\setlength{\topmargin}{-0.5cm}
\setlength{\textwidth}{15.0cm}
\setlength{\oddsidemargin}{+0.8cm}
\setlength{\topsep}{1mm}
%
% for hand-editing
%\renewcommand{\baselinestretch}{2}
%
\parskip= 4pt plus 1pt
\headsep=0.1mm
%
% -----------------------------------------------------------
%-------------------------------------------------------------------------------
\raggedbottom
\setcounter{totalnumber}{5}
\renewcommand{\textfraction}{0.1}
\renewcommand{\floatpagefraction}{0.8}
\renewcommand{\topfraction}{0.9}
\renewcommand{\bottomfraction}{0.9}
\begin{center}
  (Submitted to Physics Letters)
\end{center}
% %{Comments by {\bf June 5} to the author:}\\
% Giovanni Calderini (G.Calderini@cern.ch)\\
% {to the editor:}\\
% Fabrizio Palla (F.Palla@cern.ch)\\
% {and to the referees:}\\
% Duccio Abbaneo (D.Abbaneo@cern.ch)\\
% Roger Forty (R.Forty@cern.ch)
\newpage
%------------------------------------------------------------------------
% authors12pt.tex
%for LEP I papers only
%-----------------------------------------------------------------------
\pagestyle{empty}
\newpage
\small
%
% remember the old settings
\newlength{\saveparskip}
\newlength{\savetextheight}
\newlength{\savetopmargin}
\newlength{\savetextwidth}
\newlength{\saveoddsidemargin}
\newlength{\savetopsep}
\setlength{\saveparskip}{\parskip}
\setlength{\savetextheight}{\textheight}
\setlength{\savetopmargin}{\topmargin}
\setlength{\savetextwidth}{\textwidth}
\setlength{\saveoddsidemargin}{\oddsidemargin}
\setlength{\savetopsep}{\topsep}
%
% text dimensions for the author list
%
\setlength{\parskip}{0.0cm}
\setlength{\textheight}{25.0cm}
\setlength{\topmargin}{-1.5cm}
\setlength{\textwidth}{16 cm}
\setlength{\oddsidemargin}{-0.0cm}
\setlength{\topsep}{1mm}
\pretolerance=10000
%%%%%%%%%\begin{document}
%\centerline{EUROPEAN ORGANIZATION FOR NUCLEAR RESEARCH}
%\centerline{EUROPEAN LABORATORY FOR PARTICLE PHYSICS (CERN)}
%\vspace{1cm}
%\begin{flushright}CERN-EP-2000-   \\
%25 July 2000 - last update
%\end{flushright}
\centerline{\large\bf The ALEPH Collaboration}
\footnotesize
\vspace{0.5cm}
{\raggedbottom
\begin{sloppypar}
\samepage\noindent
R.~Barate,
D.~Decamp,
P.~Ghez,
C.~Goy,
\mbox{J.-P.~Lees},
E.~Merle,
\mbox{M.-N.~Minard},
B.~Pietrzyk
\nopagebreak
\begin{center}
\parbox{15.5cm}{\sl\samepage
Laboratoire de Physique des Particules (LAPP), IN$^{2}$P$^{3}$-CNRS,
F-74019 Annecy-le-Vieux Cedex, France}
\end{center}\end{sloppypar}
\vspace{2mm}
\begin{sloppypar}
\noindent
S.~Bravo,
M.P.~Casado,
M.~Chmeissani,
J.M.~Crespo,
E.~Fernandez,
\mbox{M.~Fernandez-Bosman},
Ll.~Garrido,$^{15}$
E.~Graug\'{e}s,
M.~Martinez,
G.~Merino,
R.~Miquel,
Ll.M.~Mir,
A.~Pacheco,
H.~Ruiz
\nopagebreak
\begin{center}
\parbox{15.5cm}{\sl\samepage
Institut de F\'{i}sica d'Altes Energies, Universitat Aut\`{o}noma
de Barcelona, E-08193 Bellaterra (Barcelona), Spain$^{7}$}
\end{center}\end{sloppypar}
\vspace{2mm}
\begin{sloppypar}
\noindent
A.~Colaleo,
D.~Creanza,
M.~de~Palma,
G.~Iaselli,
G.~Maggi,
M.~Maggi,$^{1}$
S.~Nuzzo,
A.~Ranieri,
G.~Raso,$^{23}$
F.~Ruggieri,
G.~Selvaggi,
L.~Silvestris,
P.~Tempesta,
A.~Tricomi,$^{3}$
G.~Zito
\nopagebreak
\begin{center}
\parbox{15.5cm}{\sl\samepage
Dipartimento di Fisica, INFN Sezione di Bari, I-70126
Bari, Italy}
\end{center}\end{sloppypar}
\vspace{2mm}
\begin{sloppypar}
\noindent
X.~Huang,
J.~Lin,
Q. Ouyang,
T.~Wang,
Y.~Xie,
R.~Xu,
S.~Xue,
J.~Zhang,
L.~Zhang,
W.~Zhao
\nopagebreak
\begin{center}
\parbox{15.5cm}{\sl\samepage
Institute of High Energy Physics, Academia Sinica, Beijing, The People's
Republic of China$^{8}$}
\end{center}\end{sloppypar}
\vspace{2mm}
\begin{sloppypar}
\noindent
D.~Abbaneo,
G.~Boix,$^{6}$
O.~Buchm\"uller,
M.~Cattaneo,
F.~Cerutti,
G.~Dissertori,
H.~Drevermann,
R.W.~Forty,
M.~Frank,
T.C.~Greening,
J.B.~Hansen,
J.~Harvey,
P.~Janot,
B.~Jost,
I.~Lehraus,
P.~Mato,
A.~Minten,
A.~Moutoussi,
F.~Ranjard,
L.~Rolandi,
D.~Schlatter,
M.~Schmitt,$^{20}$
O.~Schneider,$^{2}$
P.~Spagnolo,
W.~Tejessy,
F.~Teubert,
E.~Tournefier,
A.E.~Wright
\nopagebreak
\begin{center}
\parbox{15.5cm}{\sl\samepage
European Laboratory for Particle Physics (CERN), CH-1211 Geneva 23,
Switzerland}
\end{center}\end{sloppypar}
\vspace{2mm}
\begin{sloppypar}
\noindent
Z.~Ajaltouni,
F.~Badaud,
G.~Chazelle,
O.~Deschamps,
A.~Falvard,
P.~Gay,
C.~Guicheney,
P.~Henrard,
J.~Jousset,
B.~Michel,
S.~Monteil,
\mbox{J-C.~Montret},
D.~Pallin,
P.~Perret,
F.~Podlyski
\nopagebreak
\begin{center}
\parbox{15.5cm}{\sl\samepage
Laboratoire de Physique Corpusculaire, Universit\'e Blaise Pascal,
IN$^{2}$P$^{3}$-CNRS, Clermont-Ferrand, F-63177 Aubi\`{e}re, France}
\end{center}\end{sloppypar}
\vspace{2mm}
\begin{sloppypar}
\noindent
J.D.~Hansen,
J.R.~Hansen,
P.H.~Hansen,$^{1}$
B.S.~Nilsson,
A.~W\"a\"an\"anen
\begin{center}
\parbox{15.5cm}{\sl\samepage
Niels Bohr Institute, DK-2100 Copenhagen, Denmark$^{9}$}
\end{center}\end{sloppypar}
\vspace{2mm}
\begin{sloppypar}
\noindent
G.~Daskalakis,
A.~Kyriakis,
C.~Markou,
E.~Simopoulou,
A.~Vayaki
\nopagebreak
\begin{center}
\parbox{15.5cm}{\sl\samepage
Nuclear Research Center Demokritos (NRCD), GR-15310 Attiki, Greece}
\end{center}\end{sloppypar}
\vspace{2mm}
\begin{sloppypar}
\noindent
A.~Blondel,$^{12}$
G.~Bonneaud,
\mbox{J.-C.~Brient},
A.~Roug\'{e},
M.~Rumpf,
M.~Swynghedauw,
M.~Verderi,
\linebreak
H.~Videau
\nopagebreak
\begin{center}
\parbox{15.5cm}{\sl\samepage
Laboratoire de Physique Nucl\'eaire et des Hautes Energies, Ecole
Polytechnique, IN$^{2}$P$^{3}$-CNRS, \mbox{F-91128} Palaiseau Cedex, France}
\end{center}\end{sloppypar}
\vspace{2mm}
%\pagebreak 
\begin{sloppypar}
\noindent
E.~Focardi,
G.~Parrini,
K.~Zachariadou
\nopagebreak
\begin{center}
\parbox{15.5cm}{\sl\samepage
Dipartimento di Fisica, Universit\`a di Firenze, INFN Sezione di Firenze,
I-50125 Firenze, Italy}
\end{center}\end{sloppypar}
\vspace{2mm}
%\pagebreak
\begin{sloppypar}
\noindent
A.~Antonelli,
M.~Antonelli,
G.~Bencivenni,
G.~Bologna,$^{4}$
F.~Bossi,
P.~Campana,
G.~Capon,
V.~Chiarella,
P.~Laurelli,
G.~Mannocchi,$^{5}$
F.~Murtas,
G.P.~Murtas,
L.~Passalacqua,
\mbox{M.~Pepe-Altarelli}
\nopagebreak
\begin{center}
\parbox{15.5cm}{\sl\samepage
Laboratori Nazionali dell'INFN (LNF-INFN), I-00044 Frascati, Italy}
\end{center}\end{sloppypar}
\vspace{2mm}
%\pagebreak
\begin{sloppypar}
\noindent
A.W. Halley,
J.G.~Lynch,
P.~Negus,
V.~O'Shea,
C.~Raine,
\mbox{P.~Teixeira-Dias},
A.S.~Thompson
\nopagebreak
\begin{center}
\parbox{15.5cm}{\sl\samepage
Department of Physics and Astronomy, University of Glasgow, Glasgow G12
8QQ,United Kingdom$^{10}$}
\end{center}\end{sloppypar}
%\pagebreak
\vspace{2mm}
\begin{sloppypar}
\noindent
R.~Cavanaugh,
S.~Dhamotharan,
C.~Geweniger,$^{1}$
P.~Hanke,
G.~Hansper,
V.~Hepp,
E.E.~Kluge,
A.~Putzer,
J.~Sommer,
K.~Tittel,
S.~Werner,$^{19}$
M.~Wunsch$^{19}$
\nopagebreak
\begin{center}
\parbox{15.5cm}{\sl\samepage
Kirchhoff-Institut f\"r Physik, Universit\"at Heidelberg, D-69120
Heidelberg, Germany$^{16}$}
\end{center}\end{sloppypar}
\vspace{2mm}
\begin{sloppypar}
\noindent
R.~Beuselinck,
D.M.~Binnie,
W.~Cameron,
P.J.~Dornan,
M.~Girone,
N.~Marinelli,
J.K.~Sedgbeer,
J.C.~Thompson,$^{14}$
E.~Thomson$^{22}$
\nopagebreak
\begin{center}
\parbox{15.5cm}{\sl\samepage
Department of Physics, Imperial College, London SW7 2BZ,
United Kingdom$^{10}$}
\end{center}\end{sloppypar}
\vspace{2mm}
\begin{sloppypar}
\noindent
V.M.~Ghete,
P.~Girtler,
E.~Kneringer,
D.~Kuhn,
G.~Rudolph
\nopagebreak
\begin{center}
\parbox{15.5cm}{\sl\samepage
Institut f\"ur Experimentalphysik, Universit\"at Innsbruck, A-6020
Innsbruck, Austria$^{18}$}
\end{center}\end{sloppypar}
\vspace{2mm}
\begin{sloppypar}
\noindent
C.K.~Bowdery,
P.G.~Buck,
A.J.~Finch,
F.~Foster,
G.~Hughes,
R.W.L.~Jones,
N.A.~Robertson
\nopagebreak
\begin{center}
\parbox{15.5cm}{\sl\samepage
Department of Physics, University of Lancaster, Lancaster LA1 4YB,
United Kingdom$^{10}$}
\end{center}\end{sloppypar}
\vspace{2mm}
\begin{sloppypar}
\noindent
I.~Giehl,
K.~Jakobs,
K.~Kleinknecht,
G.~Quast,$^{1}$
B.~Renk,
E.~Rohne,
\mbox{H.-G.~Sander},
H.~Wachsmuth,
C.~Zeitnitz
\nopagebreak
\begin{center}
\parbox{15.5cm}{\sl\samepage
Institut f\"ur Physik, Universit\"at Mainz, D-55099 Mainz, Germany$^{16}$}
\end{center}\end{sloppypar}
\vspace{2mm}
\begin{sloppypar}
\noindent
A.~Bonissent,
J.~Carr,
P.~Coyle,
O.~Leroy,
P.~Payre,
D.~Rousseau,
M.~Talby
\nopagebreak
\begin{center}
\parbox{15.5cm}{\sl\samepage
Centre de Physique des Particules, Universit\'e de la M\'editerran\'ee,
IN$^{2}$P$^{3}$-CNRS, F-13288 Marseille, France}
\end{center}\end{sloppypar}
\vspace{2mm}
%\pagebreak %jw
\begin{sloppypar}
\noindent
M.~Aleppo,
F.~Ragusa
\nopagebreak
\begin{center}
\parbox{15.5cm}{\sl\samepage
Dipartimento di Fisica, Universit\`a di Milano e INFN Sezione di Milano,
I-20133 Milano, Italy}
\end{center}\end{sloppypar}
\vspace{2mm}
\begin{sloppypar}
\noindent
H.~Dietl,
G.~Ganis,
A.~Heister,
K.~H\"uttmann,
G.~L\"utjens,
C.~Mannert,
W.~M\"anner,
\mbox{H.-G.~Moser},
S.~Schael,
R.~Settles,$^{1}$
H.~Stenzel,
W.~Wiedenmann,
G.~Wolf
\nopagebreak
\begin{center}
\parbox{15.5cm}{\sl\samepage
Max-Planck-Institut f\"ur Physik, Werner-Heisenberg-Institut,
D-80805 M\"unchen, Germany\footnotemark[16]}
\end{center}\end{sloppypar}
\vspace{2mm}
%\pagebreak
\begin{sloppypar}
\noindent
P.~Azzurri,
J.~Boucrot,$^{1}$
O.~Callot,
S.~Chen,
A.~Cordier,
M.~Davier,
L.~Duflot,
\mbox{J.-F.~Grivaz},
Ph.~Heusse,
A.~Jacholkowska,$^{1}$
F.~Le~Diberder,
J.~Lefran\c{c}ois,
\mbox{A.-M.~Lutz},
\mbox{M.-H.~Schune},
\mbox{J.-J.~Veillet},
I.~Videau,$^{1}$
C.~Yuan,
D.~Zerwas
\nopagebreak
\begin{center}
\parbox{15.5cm}{\sl\samepage
Laboratoire de l'Acc\'el\'erateur Lin\'eaire, Universit\'e de Paris-Sud,
IN$^{2}$P$^{3}$-CNRS, F-91898 Orsay Cedex, France}
\end{center}\end{sloppypar}
\vspace{2mm}
\begin{sloppypar}
\noindent
%\samepage
G.~Bagliesi,
T.~Boccali,
G.~Calderini,
V.~Ciulli,
L.~Fo\`{a},
A.~Giassi,
F.~Ligabue,
A.~Messineo,
F.~Palla,$^{1}$
G.~Rizzo,
G.~Sanguinetti,
A.~Sciab\`a,
G.~Sguazzoni,
R.~Tenchini,$^{1}$
A.~Venturi,
P.G.~Verdini
\samepage
\begin{center}
\parbox{15.5cm}{\sl\samepage
Dipartimento di Fisica dell'Universit\`a, INFN Sezione di Pisa,
e Scuola Normale Superiore, I-56010 Pisa, Italy}
\end{center}\end{sloppypar}
\vspace{2mm}
\begin{sloppypar}
\noindent
G.A.~Blair,
G.~Cowan,
M.G.~Green,
T.~Medcalf,
J.A.~Strong,
\mbox{J.H.~von~Wimmersperg-Toeller}
\nopagebreak
\begin{center}
\parbox{15.5cm}{\sl\samepage
Department of Physics, Royal Holloway \& Bedford New College,
University of London, Surrey TW20 OEX, United Kingdom$^{10}$}
\end{center}\end{sloppypar}
\vspace{2mm}
\begin{sloppypar}
\noindent
R.W.~Clifft,
T.R.~Edgecock,
P.R.~Norton,
I.R.~Tomalin
\nopagebreak
\begin{center}
\parbox{15.5cm}{\sl\samepage
Particle Physics Dept., Rutherford Appleton Laboratory,
Chilton, Didcot, Oxon OX11 OQX, United Kingdom$^{10}$}
\end{center}\end{sloppypar}
\vspace{2mm}
%\pagebreak
\begin{sloppypar}
\noindent
\mbox{B.~Bloch-Devaux},
P.~Colas,
S.~Emery,
W.~Kozanecki,
E.~Lan\c{c}on,
\mbox{M.-C.~Lemaire},
E.~Locci,
P.~Perez,
J.~Rander,
\mbox{J.-F.~Renardy},
A.~Roussarie,
\mbox{J.-P.~Schuller},
J.~Schwindling,
A.~Trabelsi,$^{21}$
B.~Vallage
\nopagebreak
\begin{center}
\parbox{15.5cm}{\sl\samepage
CEA, DAPNIA/Service de Physique des Particules,
CE-Saclay, F-91191 Gif-sur-Yvette Cedex, France$^{17}$}
\end{center}\end{sloppypar}
%\pagebreak
\vspace{2mm}
\begin{sloppypar}
\noindent
S.N.~Black,
J.H.~Dann,
R.P.~Johnson,
H.Y.~Kim,
N.~Konstantinidis,
A.M.~Litke,
M.A. McNeil,
\linebreak
G.~Taylor
\nopagebreak
\begin{center}
\parbox{15.5cm}{\sl\samepage
Institute for Particle Physics, University of California at
Santa Cruz, Santa Cruz, CA 95064, USA$^{13}$}
\end{center}\end{sloppypar}
\vspace{2mm}
\begin{sloppypar}
\noindent
C.N.~Booth,
S.~Cartwright,
F.~Combley,
M.~Lehto,
L.F.~Thompson
\nopagebreak
\begin{center}
\parbox{15.5cm}{\sl\samepage
Department of Physics, University of Sheffield, Sheffield S3 7RH,
United Kingdom$^{10}$}
\end{center}\end{sloppypar}
\vspace{2mm}
%\pagebreak
\begin{sloppypar}
\noindent
K.~Affholderbach,
A.~B\"ohrer,
S.~Brandt,
C.~Grupen,$^{1}$
A.~Misiejuk,
G.~Prange,
U.~Sieler
\nopagebreak
\begin{center}
\parbox{15.5cm}{\sl\samepage
Fachbereich Physik, Universit\"at Siegen, D-57068 Siegen,
 Germany$^{16}$}
\end{center}\end{sloppypar}
\vspace{2mm}
\begin{sloppypar}
\noindent
G.~Giannini,
B.~Gobbo
\nopagebreak
\begin{center}
\parbox{15.5cm}{\sl\samepage
Dipartimento di Fisica, Universit\`a di Trieste e INFN Sezione di Trieste,
I-34127 Trieste, Italy}
\end{center}\end{sloppypar}
\vspace{2mm}
\begin{sloppypar}
\noindent
J.~Rothberg,
S.~Wasserbaech
\nopagebreak
\begin{center}
\parbox{15.5cm}{\sl\samepage
Experimental Elementary Particle Physics, University of Washington, Seattle, 
WA 98195 U.S.A.}
\end{center}\end{sloppypar}
\vspace{2mm}
\begin{sloppypar}
\noindent
S.R.~Armstrong,
K.~Cranmer,
P.~Elmer,
D.P.S.~Ferguson,
Y.~Gao,
S.~Gonz\'{a}lez,
O.J.~Hayes,
H.~Hu,
S.~Jin,
J.~Kile,
P.A.~McNamara III,
J.~Nielsen,
W.~Orejudos,
Y.B.~Pan,
Y.~Saadi,
I.J.~Scott,
J.~Walsh,
Sau~Lan~Wu,
X.~Wu,
G.~Zobernig
\nopagebreak
\begin{center}
\parbox{15.5cm}{\sl\samepage
Department of Physics, University of Wisconsin, Madison, WI 53706,
USA$^{11}$}
\end{center}\end{sloppypar}
}
\footnotetext[1]{Also at CERN, 1211 Geneva 23, Switzerland.}
\footnotetext[2]{Now at Universit\'e de Lausanne, 1015 Lausanne, Switzerland.}
\footnotetext[3]{Also at Dipartimento di Fisica di Catania and INFN Sezione di
 Catania, 95129 Catania, Italy.}
\footnotetext[4]{Also Istituto di Fisica Generale, Universit\`{a} di
Torino, 10125 Torino, Italy.}
\footnotetext[5]{Also Istituto di Cosmo-Geofisica del C.N.R., Torino,
Italy.}
\footnotetext[6]{Supported by the Commission of the European Communities,
contract ERBFMBICT982894.}
\footnotetext[7]{Supported by CICYT, Spain.}
\footnotetext[8]{Supported by the National Science Foundation of China.}
\footnotetext[9]{Supported by the Danish Natural Science Research Council.}
\footnotetext[10]{Supported by the UK Particle Physics and Astronomy Research
Council.}
\footnotetext[11]{Supported by the US Department of Energy, grant
DE-FG0295-ER40896.}
\footnotetext[12]{Now at Departement de Physique Corpusculaire, Universit\'e de
Gen\`eve, 1211 Gen\`eve 4, Switzerland.}
\footnotetext[13]{Supported by the US Department of Energy,
grant DE-FG03-92ER40689.}
\footnotetext[14]{Also at Rutherford Appleton Laboratory, Chilton, Didcot, UK.}
\footnotetext[15]{Permanent address: Universitat de Barcelona, 08208 Barcelona,
Spain.}
\footnotetext[16]{Supported by the Bundesministerium f\"ur Bildung,
Wissenschaft, Forschung und Technologie, Germany.}
\footnotetext[17]{Supported by the Direction des Sciences de la
Mati\`ere, C.E.A.}
\footnotetext[18]{Supported by the Austrian Ministry for Science and Transport.}
\footnotetext[19]{Now at SAP AG, 69185 Walldorf, Germany.}
\footnotetext[20]{Now at Harvard University, Cambridge, MA 02138, U.S.A.}
\footnotetext[21]{Now at D\'epartement de Physique, Facult\'e des Sciences de Tunis, 1060 Le Belv\'ed\`ere, Tunisia.}
\footnotetext[22]{Now at Department of Physics, Ohio State University, Columbus, OH 43210-1106, U.S.A.}
\footnotetext[23]{Also at Dipartimento di Fisica e Tecnologie Relative, Universit\`a di Palermo, Palermo, Italy.}
% restore the previous settings
%
\setlength{\parskip}{\saveparskip}
\setlength{\textheight}{\savetextheight}
\setlength{\topmargin}{\savetopmargin}
\setlength{\textwidth}{\savetextwidth}
\setlength{\oddsidemargin}{\saveoddsidemargin}
\setlength{\topsep}{\savetopsep}
%%%%%%%%%%%%%%%%%%%%%%%%%%%%%%%%%%%%%%%%%
\normalsize
\newpage
\pagestyle{plain}
\setcounter{page}{1}

%
%------------------------------------------------------------------------
\setcounter{page}{1}
 
\section{Introduction}
 
Measurements of the individual $b$ hadron lifetimes represent an important 
test of the present knowledge of nonspectator effects in the $b$ hadron
decay dynamics, such as Pauli interference, $W$ exchange and weak annihilation.
Based on the heavy quark expansion formalism, the difference between
the lifetimes of the $b$ baryons and mesons is predicted to depend on $1/m_b^2$ and higher order terms, whereas meson-meson differences depend only on $1/m_b^3$ and
higher order terms \cite{teorici}. The predicted hierarchy is $\tLamb<\t0 \sim\tBs<\tp$. Differences are expected to be at the level of a few percent, which sets the scale of the experimental precision required.

This paper reports an improved   measurement of the \b0 \ and \bp\ lifetimes with the \aleph\ detector at LEP, using approximately four million hadronic decays of the Z, collected in the period 1991--1995. This data sample was recently reprocessed, achieving higher efficiency and better resolution in the track reconstruction, which is highly beneficial for this analysis. 

Semileptonic decays of \b0 \ and \bp\ mesons are
partially reconstructed by identifying events containing a lepton
(electron or muon) with an associated \d0 \ or \ds\ meson. The resulting
\d0 -lepton (\dl) and \ds-lepton
(\dsl) event samples consist mostly of \bp\ and \b0 \ decays,
respectively (charge conjugate modes are implied throughout
this paper).

Previous measurements of the  \b0 \ and \bp\ lifetimes are reported in
\cite{ref:bl95,OTHER}.

\section{The ALEPH detector}
\label{sec:aleph}
 
A detailed description of the \A\ detector can be found 
elsewhere~\cite{detect,vdetdescr}.
A high resolution
vertex detector (VDET) consisting of two layers of
silicon with double-sided readout
provides measurements in the $r\phi$ and $z$ directions
at average radii of 6.5 cm and 11.3 cm, with
$ 12$ \micron\ precision at normal incidence. The VDET provides full azimuthal
coverage, and polar angle coverage to $|{\cos\theta}|<0.85$
for the inner layer only and $|{\cos\theta}|~<~0.69$ for both
layers.
Outside the VDET, particles
traverse the inner tracking chamber (ITC) and the
time projection chamber (TPC).
The ITC is a cylindrical drift chamber with eight axial
wire layers at radii of 16 to 26 cm.
The TPC measures up to 21 space points
per track at radii between 40 and 171 cm, and
also provides a measurement of the specific 
ionization energy loss (\dedx) of each charged track.
These three detectors form the tracking system, which is immersed in 
a 1.5 T axial magnetic field provided by a superconducting solenoid.
The combined tracking system yields a momentum resolution transverse to the beam axis of 
$ \sigma(p_T)/p_T = 6 \times 10^{-4} \ p_T \oplus 0.005 $ ($p_T$ {\rm in \gevc}).
The resolution of the three-dimensional impact parameter 
for tracks having two VDET hits can be 
parametrized as $\sigma=25 ~\micron +95 ~\micron/p$ ($p$ in \gevc)~\cite{ref:performance}.

The electromagnetic calorimeter (ECAL) is a
lead/wire-chamber sandwich operated in proportional mode.
The calorimeter is read out in projective towers that
subtend typically $0.9^\circ\times 0.9^\circ$
in solid angle, segmented in three longitudinal sections.
The hadron calorimeter (HCAL) uses the iron return yoke as absorber.
Hadronic showers are sampled by 23 planes of streamer tubes,
with analogue projective tower and digital hit pattern readout.
The HCAL is used in combination with two double layers
of muon chambers outside the magnet for muon identification.

Recently the LEP1 data were reprocessed using improved reconstruction 
algorithms. The features that are particularly relevant for the enhancement of the charmed meson reconstruction efficiency are the following. A new VDET pattern recognition algorithm allows groups of several
nearby tracks which may share common hits to be analysed together,
to find the hit assignments that minimize the overall $\chi^2$ for the
event. The improvement in the hit association
 efficiency is more than 2\% (from 89.2\% to 91.0\% in  $r\phi$ and from 85.6\% to 88.2\% in $z$).
Information on the drift time from the TPC wires is combined with that
obtained from the pads to reduce the error in the $z$ coordinate by 
a factor of two. A 30\% improvement in the $r\phi$ coordinate
resolution is achieved for low momentum tracks by correcting the 
pad coordinates for ionisation fluctuations along the tracks as 
measured by the wires. The particle identification ($dE/dx$) is 
improved by combining pulse height data from the TPC pads with that
of the wires. The improvements in the charmed meson reconstruction efficiencies with 
respect to the previous analysis range from 10 to 30\%, depending on the
decay channel.

\section{Event selection\label{sec:sel}}
The \dsl\ and \dl\ event samples consist of an identified lepton 
($e$ or $\mu$) associated with a fully reconstructed \ds\ or \dz\ candidate. 
The selection of muons and electrons is described in detail in
\cite{ref:leptons}. For this analysis, lepton candidates are
required to have a momentum of at least 2.0~\gevc\ for electrons
and 2.5~\gevc\ for muons.
 
The \ds\ and  \d0 \ candidates are reconstructed from charged
tracks and $\pi^0$'s that form an angle of less than
$45^\circ$ with the lepton candidate.
These charged tracks are also required to
intersect a cylinder of radius 2 cm
and half-length 4 cm centered on the nominal interaction point,
to have at least 4 hits in the TPC, a polar
angle $\theta$ such that $|{\cos \theta}|<0.95$ and a transverse 
momentum greater than 200~\mevc.

Photons and $\pi^0$'s are identified in the ECAL. The four-momenta
of $\pi^0$ candidates are computed by adding the photon momenta
when the $\gamma\gamma$ invariant mass is consistent with the $\pi^0$ mass.
The energy  of the $\pi^0$ is then recomputed using the kinematical
constraint of the  $\pi^0$ mass \cite{ref:performance}. The energy resolution achieved is about 6.5\%, almost independent of the energy. For
reconstructing the $D^0$ candidates, only $\pi^0$'s with momenta greater than 2 \gevc\ are used.

For all charged kaons used to reconstruct the $D^0$ it is required that the $dE/dx$ be within $3\sigma$ of that expected from a kaon, except for the \dsdpi, \dkpi decay channel. The kaon has the same sign as the lepton coming from the $B$ semileptonic decay, therefore this charge correlation is required.

$K^0_{\rm S}$ candidates are reconstructed from pairs of oppositely charged tracks. The two tracks are required to be inconsistent with
originating from the interaction point and the $K^0_{\rm S}$ candidate is rejected
if the measured mass is more than $2\sigma$ ($\pm 10\;$\mevcc) from the
nominal $K^0_{\rm S}$ mass.

Tracks coming from the  $D^0$ decays are required to form a common vertex with 
a $\chi^2$ confidence level greater than 1\%. 
In  the $D^0$ decay channels that do not contain  a $\pi^0$ in the
final state, if more than one combination satisfies the selection criteria, the one with the smallest \chisq\ of the $D$ vertex fit is selected.
For the decay modes where the multiple candidates originate from different  $\pi^0$ combinations, for a given detected lepton, the candidate with  \delm\ closest to the nominal value is chosen in case of $D^\star$ selection, otherwise the highest momentum $D^0$ candidate is chosen.

In order to reject the background from charm, to
improve the signal to background ratio and to ensure well-measured
decay lengths, additional selection criteria are applied to all the subsamples.
The invariant mass of the $\dords\ell$ system is required to be greater than 3~\gevcc, where \dords\ can be \ds\ or \d0 .
This cut significantly reduces the charm background
while keeping $\sim 85\%$ of the signal.
To exploit the high precision of the silicon vertex detector,
the lepton track and at least two tracks from the
\d0 \ decay are required to have at least one VDET
hit in both the \rphi\ and $z$ projections.

\subsection{The {\boldmath $D^\star$}-lepton selection}
\ds\ candidates are identified via the decay \dsdpi, followed 
by \dkpi, \dkppp, \dkppz\ or \dkpps. 
The difference in mass between the \ds\
and \d0 \ candidates must lie within 1.5~\mevcc\
(approximately two standard deviations of the experimental resolution)
of the nominal value of 145.4~\mevcc.

%\paragraph{\boldmath{$D^0 \rightarrow K^-\pi^+$} channel.}

In the $D^0 \rightarrow K^-\pi^+$ channel, the \d0 \ momentum $p_{D^0}$ is required to be greater than 5~\gevc. 

For the $D^0 \rightarrow K^-\pi^+\pi^-\pi^+$ channel $p_{D^0}$ must be greater than 8~\gevc, and at least two of the \d0 \ decay tracks must have $p>1$~\gevc. 

%\paragraph{\boldmath{$D^0 \rightarrow K^-\pi^+\pi^-\pi^+$} channel.} 
%A cut $p_{D^0} > 8$~\gevc\ is applied. 
%At least two of the \d0 \ decay tracks must have $p>1$~\gevc. 
%\paragraph{\boldmath{$D^0 \rightarrow K^-\pi^+\pi^0$} channel.}
In the case of the $D^0 \rightarrow K^-\pi^+\pi^0$ channel,
the momentum of the reconstructed \d0 \ is required to be greater than
10 \gevc, and the two charged tracks in the decay are required to
have $p>$ 0.5 \gevc.
Futhermore the decay kinematics are required to be consistent with 
one of the three resonant decays:
$\d0 \rightarrow K^- \rho^+$,~$\d0 \rightarrow K^{\star-} \pi^+ $,
~$\d0 \rightarrow {\bar K}^{\star0} \pi^0$. 
For each decay hypothesis the mass of the resonant particle and the
helicity angle $\theta_H$ are calculated\rlap.\footnote{ The helicity angle $\theta_H$ is defined as the angle between the scalar particle and one of the decay products of the vector particle, calculated in the rest frame of the  vector particle. } 
If the mass is consistent with the nominal resonance mass value, within
its natural width, and if
$|\cos \theta_{H}| > 0.4$, the candidate is considered to be 
consistent with the resonant decay hypothesis. For these resonant decays 
a $\cos^2 \theta_{H}$ distribution is expected.

%\paragraph{\boldmath{$D^0 \rightarrow K^0_{\rm S}\pi^+\pi^-$} channel.}
Finally, in the $D^0 \rightarrow K^0_{\rm S}\pi^+\pi^-$ channel, the momentum of the reconstructed $D^0$ is required 
to be greater than 4 \gevc.
The same technique as in the \dkppz\ channel is used to tag the $K^{\star-}$
resonance. The sign of the resonance is used to distinguish between
$D^0$ and $\bar{D}^0$.

\subsection{The \boldmath{$D^0$}-lepton selection}
The \dl\ sample consists of events with a lepton and a \d0 \
candidate, where the \d0 \ is not the decay product of a reconstructed \ds.
The \d0 \ candidates are identified using the same decay modes as for the \dsl sample.
For this sample, the background is larger because the 
\ds--\d0 \ mass difference criterion is not applicable.

For all the decay modes selected in the \dl sample, a search for the additional
pion is performed
to reject \d0 \ candidates coming from \dsdpi. If a pion candidate  is found yielding a \ds--\d0 \
mass difference within 6~\mevcc\ of the nominal value,
the \dl\ candidate is rejected. 
The efficiency for reconstructing the additional pion and rejecting \d0 's coming from \ds\ decays is found to be 86\% from the Monte Carlo simulation.

%\paragraph{\boldmath{$D^0 \rightarrow K^-\pi^+$} channel.} 
In the $D^0 \rightarrow K^-\pi^+$ channel, the reconstructed \d0 's are required to have ${\mbox p_{D^0}>8~\gevc}$, $p_K>2$~\gevc\ and  $p_\pi>1.5$~\gevc.
%\paragraph{\boldmath{$D^0 \rightarrow K^-\pi^+\pi^-\pi^+$} channel.} 
In the $D^0 \rightarrow K^-\pi^+\pi^-\pi^+$ channel
the  $D^0$ momentum is required to exceed $ 12$~\gevc\, while the
kaon and the three pions must have momenta greater than 2 and 1 \gevc, respectively.

%\paragraph{\boldmath{$D^0 \rightarrow K^-\pi^+\pi^0$} channel.} 
The cuts on the kaon and pion momenta are tightened for the $D^0 \rightarrow K^-\pi^+\pi^0$  channel, to
$p_K>3$~\gevc\ and  $p_\pi>2$~\gevc, while the reconstructed \d0 \ 
must have ${\mbox p_{D^0}>12~\gevc.}$
The same cuts on the three resonances are used 
as in the \dsl event sample.

%\paragraph{\boldmath{$D^0 \rightarrow K^0_{\rm S}\pi^+\pi^-$} channel.} 
Finally, for the $D^0 \rightarrow K^0_{\rm S}\pi^+\pi^-$  channel  a cut is 
applied of at least $1.5$~\gevc\ 
on the kaon momentum and $1.0$~\gevc\ on the two pion momenta. The momentum of the reconstructed \d0 \ candidates is required to
be greater than 10 \gevc. 
The same technique is used
to tag the $K^{\star-}$ resonance as in the \dsl \ event sample.

\subsection{\boldmath{$B$} meson reconstruction}

The $B$ decay vertex position is estimated by vertexing the reconstructed \d0 \ track with the lepton.
Events are rejected if the $B$ vertex fit gives a $\chi^2$ probability 
less than 1\%.

\begin{figure}
\begin{center}
\mbox{\psfig{file=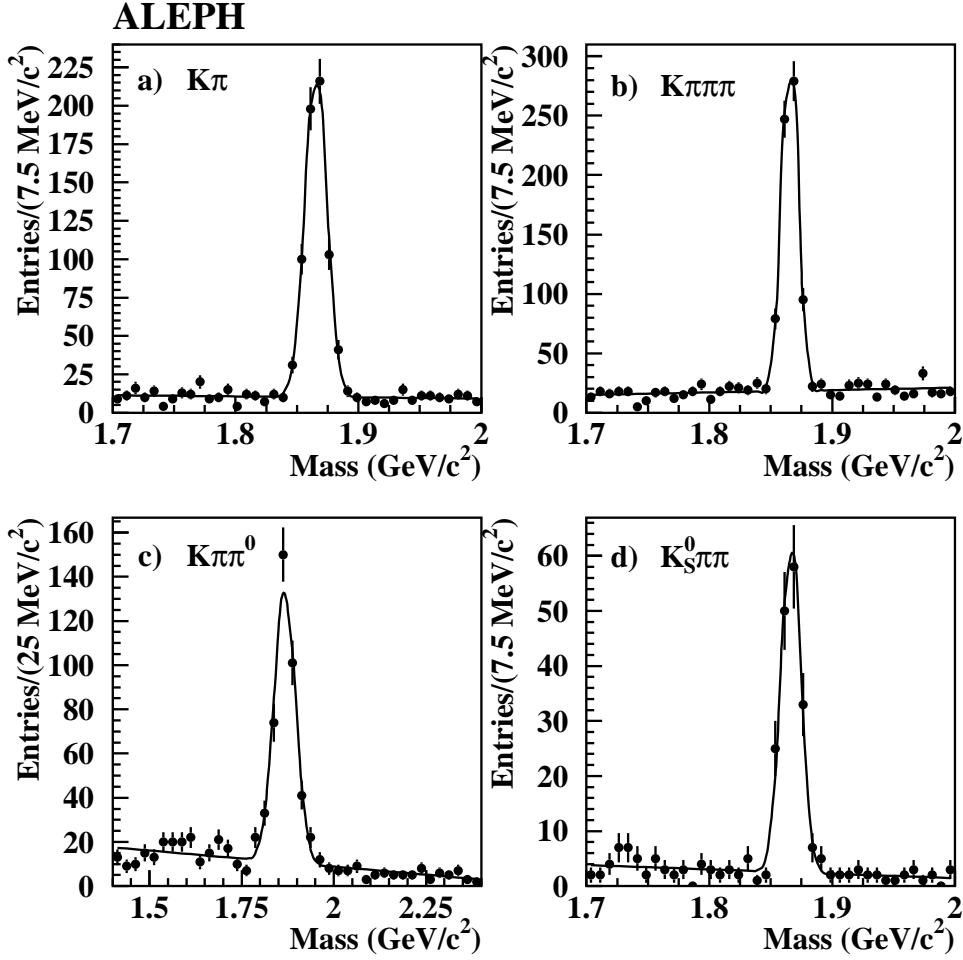,width=14cm}}
\end{center}
\caption{\small The invariant mass of \d0 \ candidates for the four subsamples
in the \dsl \ event selection:
a) $D^0\rightarrow K^-\pi^+$, b) $D^0\rightarrow K^-\pi^+
\pi^-\pi^+$,
c) $D^0 \rightarrow K^-\pi^+\pi^0$ (notice the different mass scale),
d) $D^0\rightarrow K^0_{\rm S}\pi^-\pi^+$. The  superimposed curves are the results of the fit described in the text.}
\label{mass}
\end{figure}

\begin{figure}
\begin{center}
\mbox{\psfig{file=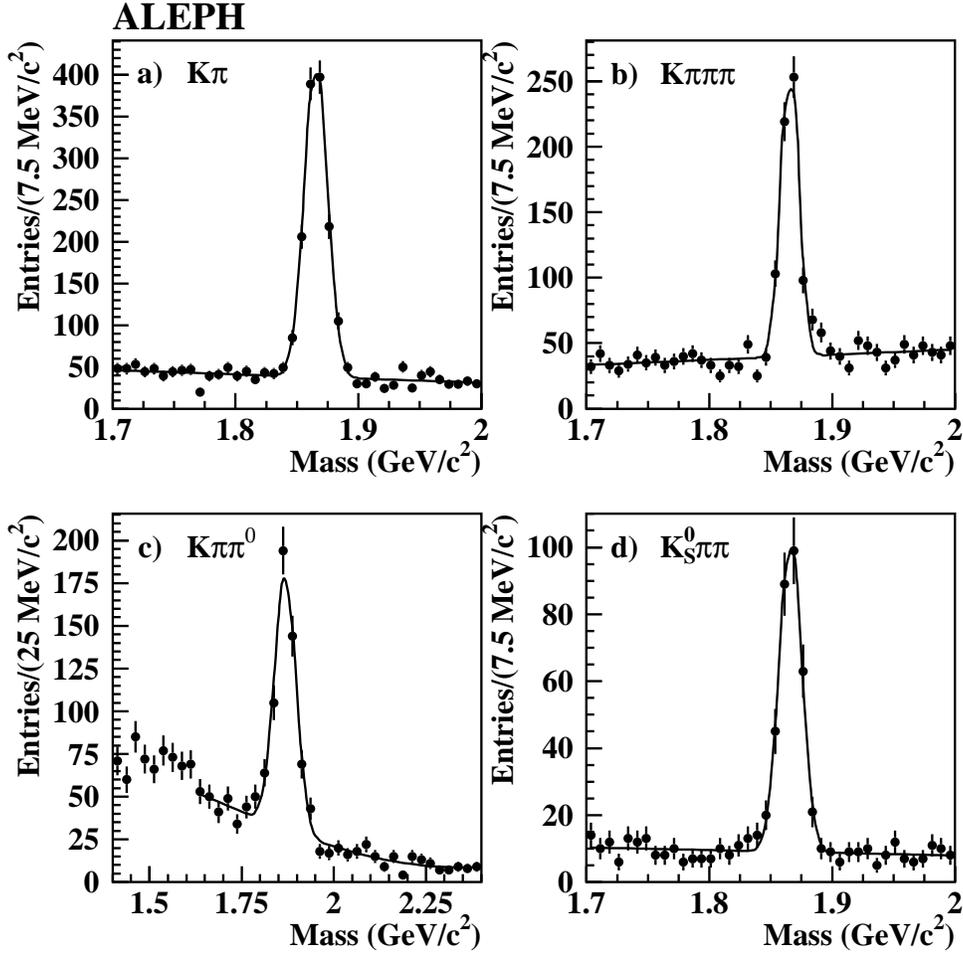,width=14cm}}
\end{center}
\caption{\small The invariant mass of \d0 \ candidates for the four subsamples
in the \dl \ event selection:
a) $D^0\rightarrow K^-\pi^+$, b) $D^0\rightarrow K^-\pi^+
\pi^-\pi^+$, c) $D^0\rightarrow K^-\pi^+\pi^0$ (notice the different mass scale),
d) $D^0 \rightarrow K^0_{\rm S}\pi^+\pi^-$. The superimposed curves are the results of the fit described in the text.}
\label{mass2}
\end{figure}

The \d0 \ candidate mass spectra for the four subsamples 
in the \dsl\ event selection are shown in Fig.~\ref{mass}, and in the 
\dl \ event selection in Fig.~\ref{mass2}.
The fitted curves consist of a Gaussian function for the
signal plus a linear background.
For the $D^0 \rightarrow K^-\pi^+\pi^0$ in the \dl sample a Gaussian tail is used to describe the background;  masses 
below $\mbox{{\rm 1.65 GeV/}$c^2$}$ are excluded to avoid the broad enhancement due to 
the missing $\pi^0$ in the $D^0\to K^-\pi^+\pi^0\pi^0$ decay. The fitted \d0 \ width and
the fitted number of signal and background events
within a window of $\pm 2\sigma$ around the fitted mass
are shown in Table~\ref{tab:mass} for the different samples.

Events reconstructed within two standard deviations
of the fitted  \d0 \ mass are selected for the lifetime analysis,
resulting in 1880 \dsl\ and 2856 \dl\ candidates.
The decay length is calculated for these events by
reconstructing the primary and $B$ decay vertices in three
dimensions. 
The primary vertex reconstruction algorithm~\cite{ref:gammabb} 
applied to simulated \bbbar\ events  yields an average resolution
of $50 ~\micron\times 10 ~\micron\times 60 ~\micron$ (horizontal $\times$
vertical $\times$ beam direction).

The distance between the primary and $B$ decay vertices is projected onto
the direction defined by the momentum of the $\dords\ell$ system.
The uncertainty on the flight direction due to the missing neutrino
induces a negligible error on the decay length.
The resolution on the $B$ decay length
is on average 250 $\micron$, compared with an average $B$ decay
length of about $2.5$ mm.

 For the \dkppz\ channel, the \pz\ momentum is included when extrapolating the neutral \d0 \ track backwards to form the $B$ vertex.
In the case of \dsl\ events, the soft pion from the \ds\ decay
does not improve the resolution on the $B$ decay length and
is therefore not used in the reconstruction of the $B$ vertex. 

%\newpage
Because the selected decays contain an undetected neutrino, 
the $B$ momentum is not known precisely and is reconstructed using 
an energy  flow technique as described in~\cite{ref:frag}. 
A further correction is applied by evaluating for Monte Carlo events 
the $\kappa$ distribution, defined as
\be \kappa = \frac{(\beta\gamma)_{\rm reco}}{(\beta\gamma)_B}. \ee
Because this distribution depends on the selection criteria applied,
separate $\kappa$ distributions are calculated for each subsample.
The momentum resolution obtained is between 8 and 10\%, depending on the decay 
channel. 

\begin{table}
\caption{\small Fitted \d0 \ width ($\sigma$), number of \d0 \ candidates and fraction of
background events falling within a mass window of $\pm 2\sigma$.
The uncertainties are statistical only.}
\label{tab:mass}
\begin{center}
\begin{tabular}{|ll|c|c|c|} \hline
% \multicolumn{2}{|c|}{Subsample}
% & Width (\mevcc)& Signal events & Background events \\  \hline
% \dsl &  \dkpi & $8.6\pm0.3$ & $607\pm25$ & $43\pm7$ \\
%      & \dkppp & $6.6\pm0.3$ & $606\pm26$ & $64\pm8$ \\ 
%      & \dkppz & $30.3\pm1.6$ & $344\pm20$ & $50\pm8$ \\
%      & \dkpps & $8.3\pm0.7$ & $155\pm12$ & $10\pm3$ \\ \hline
% \dl  &  \dkpi & $8.7\pm0.3$ & $1138\pm37$ & $174\pm13$ \\ 
%      & \dkppp & $7.2\pm0.5$ & $510\pm27$ & $154\pm13$ \\ 
%      & \dkppz & $28.4\pm1.9$ & $418\pm27$ & $145\pm13$ \\ 
%      & \dkpps & $8.6\pm0.8$ & $273\pm15$ & $44\pm6$ \\ \hline
 \multicolumn{2}{|c|}{Subsample}                              
& Width (\mevcc)& Candidate events & Background fraction \\  \hline
 \dsl &  \dkpi & $8.6\pm0.3$ & $651$ &  $0.066\pm0.004 $ \\       
      & \dkppp & $6.6\pm0.3$ & $670$ &  $0.096\pm0.004$ \\       
      & \dkppz & $30.3\pm1.6$ &$394$ & $0.127\pm0.008$ \\      
      & \dkpps & $8.3\pm0.7$ & $165$ &  $0.061 \pm0.006$ \\ \hline
 \dl  &  \dkpi & $8.7\pm0.3$ &$1312$ & $0.133\pm0.006$ \\    
      & \dkppp & $7.2\pm0.5$ & $664$ &  $0.232\pm0.012$ \\     
      & \dkppz & $28.4\pm1.9$ &$563$ & $0.258\pm0.012$ \\    
      & \dkpps & $8.6\pm0.8$ & $317$ &  $0.139\pm0.009$ \\ \hline
\end{tabular}
\end{center}
\end{table}

\section{Lifetime measurement}
\label{sec:bg}
An unbinned likelihood fit for the lifetimes is performed.
For each event the
probability of observing a proper time $t$ given the lifetime $\tau$ is
calculated:
the probability density function $F(t,\sigma_t,\tau)$
is obtained by convoluting an exponential
distribution with the properly normalised $\kappa$ distribution and
with a Gaussian function which takes into account the resolution on the
decay length.

Both the \dsl\ and \dl\ samples contain a mixture
of \b0 \ and \bp\ decays and the \bp/\b0 \ mixture in the samples depends
on the ratio of the lifetimes, as  discussed in Section \ref{samplecomp}.
Therefore, to measure the \b0 \ and \bp\ lifetimes a simultaneous maximum
likelihood fit is performed to all the events.
The likelihood function
contains three components for each sample and is written as
\begin{eqnarray}
{\cal L} & = & \prod_{i=1}^{N_{D^{\star}\ell}}
 \left[ \fsp({{\tp}/{\t0 }}) F(t_i,\sigma_i,\tp) + 
 \fso({{\tp}/{\t0 }}) F(t_i,\sigma_i,\t0 ) +
 \fsbg F_{\rm bkg}^\star(t_i)  \right]\nonumber \\
         &   \times & \prod_{i=1}^{N_{D^{0}\ell}}
 \left[ \fop({{\tp}/{\t0 }}) F(t_i,\sigma_i,\tp) + 
 \foo({{\tp}/{\t0 }}) F(t_i,\sigma_i,\t0 ) +
                  \fobg F_{\rm bkg}^0(t_i)\right]. \label{eq:like1}
\end{eqnarray}
The coefficients \fsp\ and \fso\ are the fractions of the \dsl\ sample
arising from \bp\ and \b0 \ decays, respectively. Similarly,
\fop\ and \foo\ are the fractions of \bp\ and \b0 \ decays 
in the \dl\ sample.
The coefficients \fsbg\ and \fobg\ are the background fractions of
the samples, while
the functions $F_{\rm bkg}^{\star}(t)$ and $F_{\rm bkg}^0(t)$
are their normalised proper time distributions.
 
\subsection{Backgrounds}
Background contamination arises from the following sources:
\begin{enumerate}
  \item[(1)] combinatorial background, i.e.~candidates with a fake
  \dords;
  \item[(2)] the process $\bar{B}\rightarrow D_s^- D^{(*)}X$, followed
      by $D_s^- \rightarrow \ell^-X$, giving rise to a real \dords\
      and a real lepton;
  \item[(3)] a real \dords\ meson accompanied by a fake or nonprompt
  lepton, from
   $Z\rightarrow b{\bar b}$ or $Z\rightarrow c{\bar c}$ events.
\end{enumerate}
 
Source (1) is the dominant background. Its
contribution is determined from
a fit to the \d0 \ mass distributions, and its magnitude is given in
Table~\ref{tab:mass} for the various subsamples. The proper time
distribution for this source is determined
from the data by selecting events from the high mass sideband of the \d0 \ peak.
The same selection criteria described in Section~\ref{sec:sel}
are applied to the background
samples, except that the requirement on the \ds--\d0 \ mass difference
in the case of the \dsl\ events is removed to increase
the statistics.
A function consisting of a Gaussian plus positive and negative exponential
tails is used to describe these data.
 
The contribution from source (2) is calculated from the
measured branching ratios for this process~\cite{ref:dsd} and 
a Monte Carlo simulation to determine the detection efficiency;
it accounts for a contamination which is
about 2--3\% of the sample, depending on the channel.

The background from source (3) is estimated from the measured
hadron-lepton misidentification probabilities~\cite{ref:leptons} and the
measured inclusive \d0 \ and \ds\ rates. An independent estimate is
obtained using wrong-sign ($\ds\ell^+$ or $\d0 \ell^+$) events, and 
is found to be consistent. This background
source contributes between $2$ and $5\%$  of the sample, depending on the channel.
To characterize the proper time distribution of this background, it is
further subdivided into three distinct components: fake leptons coming from the primary vertex in \ccbar\ and \bbbar\ events, and fake leptons coming from a
decaying $b$ hadron.

The proper time distribution for sources (2) and (3) are determined from
simulated events.

\subsection{Sample compositions}
\label{samplecomp}
Both the \dsl\ and \dl\ samples contain a mixture
of \b0 \ and \bp\ decays. In order to estimate the cross contamination, the
individual semileptonic branching ratios of the \b0 \ and \bp\ must be 
determined. 
%  Each of the two event samples contains a mixture of \b0 \ and \bp\
% and the sample compositions
% (the coefficients \fsp, \fso, \fop\ and \foo\ of Eq.~\ref{eq:like1})
% must be calculated to
% complete the specification of the likelihood function.
The evaluation follows the same procedure as given in the appendix of \cite{ref:bl95}.
An important input in this evaluation is the knowledge of the branching
ratios for the decay modes $B \rarr D^{\star\star} \pi \ell \nu$, where 
$D^{\star\star}$ represents any of the ${\mbox p{\rm -wave}}$ resonances as well as the nonresonant $D^{(\star)}\pi$ states.
 The most recent ALEPH and DELPHI results~\cite{ref:dss,ref:delphi_new_d} for both
the resonant and the nonresonant components are used in the calculation,
leading to  a significant reduction in the resulting uncertainty compared to 
the  previous analysis of \cite{ref:bl95}.
%
%\begin{eqnarray} 
%&& BR\;(\overline{B}^0\;\rightarrow D^0\pi^+{\ell^-} \overline{\nu})
%\;+\;BR\;(\overline{B}^0\;\rightarrow D^{\star0}\pi^+{\ell^-} \overline{\nu}) = 
%0.0123 \pm  0.0025 \\
%&& BR\;({B}^-\;\rightarrow D^+\pi^-{\ell^-} \overline{\nu})\;+\;
%BR\;({B}^-\;\rightarrow D^{\star+}\pi^-{\ell^-} \overline{\nu}) 
%= 0.0134\pm0.0034 \\
%&& BR\;({B}^-\;\rightarrow D^{\star+}\pi^-{\ell^-} \overline{\nu}) = 0.0125\pm0.0032 
%\end{eqnarray}
The \b0 \ and \bp\ content of the two samples
are calculated using  as input the measured values of the branching ratios
given in Table~\ref{tab:b0br}. 
%\cite{ref:dss,pdg98}.
% An additional constraint based on isospin conservation is applied, as well
% as the condition that the sum of the exclusive channels is equal to the 
% inclusive \b0 \ semileptonic branching ratio.
% The \bp\ branching fractions are computed from
% \be {\rm BR}(\bm\rightarrow \ell^-X)=
%     \frac{\tp}{\t0 } {\rm BR}(\bb\rightarrow\ell^-X^\prime),  \label{eq:bp}\ee
% which follows from the expectation that
% the partial semileptonic decay widths are equal.
The sample composition
is then calculated by considering the \b0 \ and \bp\
decay channels that contribute to the \dsl\ and \dl\ samples~\cite{ref:bl95,ref:blgls}, taking into account  the probability of $0.147\pm0.015$ that a \dsl\ event is mistakenly reconstructed as a \dl\ event.

As a consequence of this procedure,
the coefficients \fsp, \fso, \fop\ and \foo\
appearing in the likelihood function
(Eq.~\ref{eq:like1}) depend on the lifetime ratio.
For equal lifetimes,  about $85\%$ of the $B$ decays in the \dsl\ sample are attributed to \b0 ,
while about $80\%$ of the \dl\ sample $B$ decays come from \bp.
 
\subsection{Fit results}
The fit to the proper time distributions of the
\dsl\ and \dl\ events
is performed to determine the two free parameters
\t0 \ and \tp. The values obtained are
\begin{eqnarray*}
       \t0 & = & 1.518 \pm 0.053 \ \mathrm{ps,}       \\
       \tp & = & 1.648 \pm 0.049 \ \mathrm{ps,}
\end{eqnarray*}
where the errors are statistical only. The statistical correlation coefficient is
$-0.35$. The ratio of the lifetimes is
$$   \tp/\t0  =  1.085 \pm 0.059, $$
taking into account the correlation.

\begin{figure}
\begin{center}
\begin{tabular}{c}
\mbox{\psfig{file=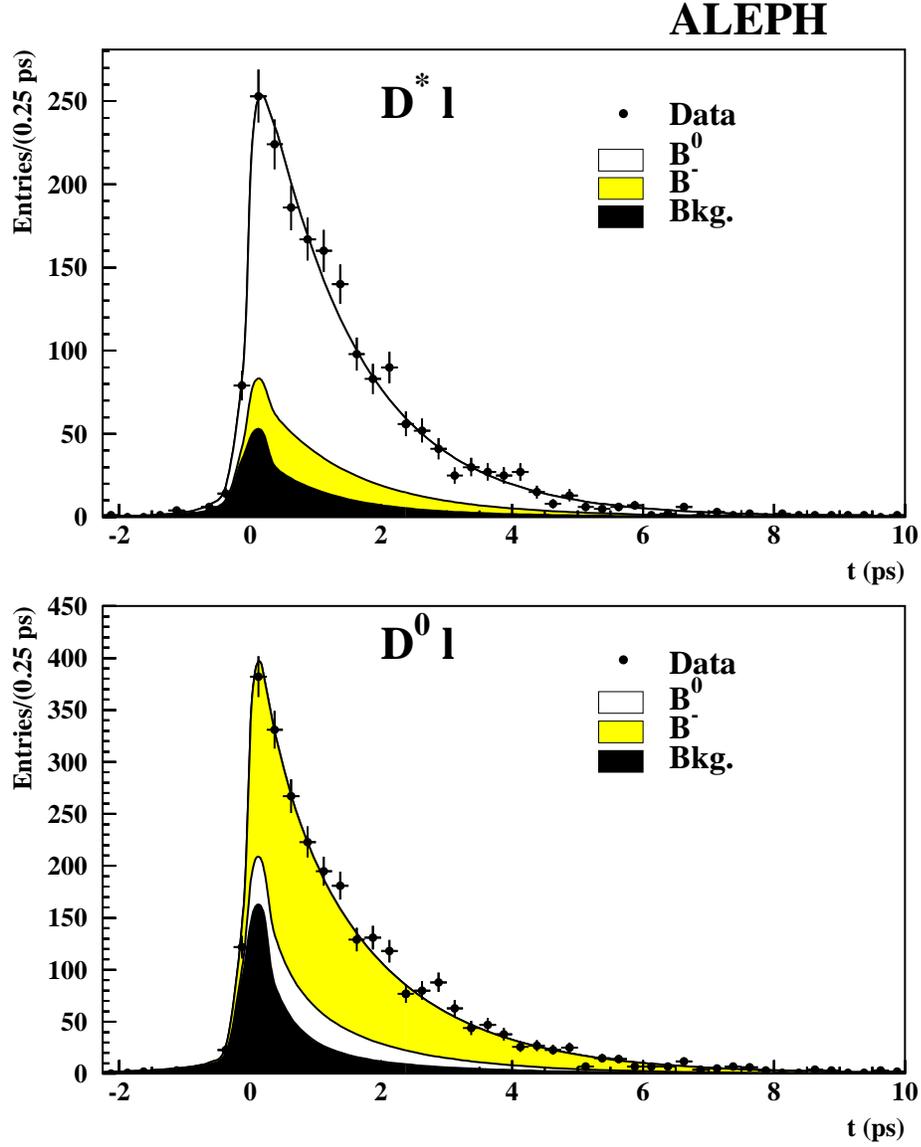,height=16.0cm}}
\end{tabular}
\end{center}
\caption{\small Proper time distributions with the result of the fit overlaid 
for the two samples. The plots show the background 
contributions to the samples, together with the \b0 \ and \bp \ components.
}
\label{fit1}
\end{figure}

The proper time
distributions for the two samples are shown in Fig.~\ref{fit1},
with the results of the fit overlaid.
%
% 
%As a check on the procedure, a measurement of the \d0 \ lifetime
%has been performed.
%The \d0 \ flight distance is
%calculated as the distance between the $B$ and \d0 \ decay vertices, projected
%onto the \d0 \ direction. 
%An unbinned likelihood fit to the 3920 events yields
%$$ \tdo = 0.404\pm0.012\mathrm{\ (stat)\ ps.}$$
%in agreement with the world average value
%$\tdo = 0.415\pm0.004$~ps~\cite{pdg98}.
%
\subsection{Systematic uncertainties\label{sec:syst}}
%The possible sources of systematic uncertainty are considered here 
%and are summarized in Table~\ref{tab:syst}.
%
%The uncertainty in the $B$ momentum reconstruction is due mainly to the
%$\kappa$-distribution used for the final $B$ momentum correction.
%The effects which can modify the $\kappa$ function, with a consequence
%on the extracted lifetimes are studied. 
%The first one is due to the $D^{\star*}$ content in semileptonic $B$ decays, which
%determines a narrower $\kappa$ distribution, due to the smaller available phase-space. The reconstructed $D^{(*)}$-lepton combination presents also
%a softer momentum spectrum. The fraction of $D^{\star*}$ decays is varied within
%the uncertainty on the branching ratios in the Monte Carlo samples from 
%which the $\kappa$-distributions are extracted. A second effect which
%is studied is the dependence of the $\kappa$-distribution on the particular 
%cuts used in the analysis. This effect, together with the uncertainty
%on the $B$ fragmentation function in the simulation, gives approximately
%a $2\%$ variation in the $B$ momentum determination. Since the $\kappa$
%distribution acts in the same way on $\tau_-$~and $\tau_0$, this systematic cancels
%out in the lifetime ratio determination.
The sources of systematic uncertainty are discussed in the following,
and the estimated errors are summarized in Table~\ref{tab:syst}.

The uncertainty in the $B$ momentum reconstruction is dominated by the 
uncertainty in the $\kappa$ distribution. The effects that can modify the $\kappa$ function with a consequence
on the extracted lifetimes are studied. The first is due to the $D^{(\star)}\pi$ content
of semileptonic $B$ decays, which affects the $\kappa$ function due to the smaller
phase space available and the softer momentum spectrum
of the reconstructed $D^{(\star)}\ell$ system. The fractions of $D^{(\star)}\pi$
decays are varied in the simulation within experimental errors.
Other effects studied are the dependence of the $\kappa$ distribution on the analysis cuts and the $b$ fragmentation function. These two effects 
combined give a relative uncertainty of about $2\%$ in the $B$ momentum determination.
Uncertainties in the $\kappa$ distribution propagate in the same way to
\tp\ and \t0 , and therefore have a small effect in the ratio.

Uncertainties in the background fractions and proper time
distributions are considered. 
% The combinatoric background level is estimated
% on in the data by a fit to the $D^0$ mass spectra for the different samples. 
The combinatorial background, estimated in the data  from fits to the $D^0$ 
mass spectra, is reported in Table \ref{tab:mass} for each individual
channel. The systematic uncertainty due to this source is estimated by varying
the combinatorial background in the fit for each channel within its statistical errors  in turn and taking the sum in quadrature.
The contamination from physics background is evaluated using the simulation, by varying the
fraction within the estimated uncertainty and repeating the fits. The uncertainty 
on the total background level is calculated by combining the two above sources.
%  The corresponding background fraction is varied within its uncertainty to 
%study the effect on the \tp ~and \t0 ~fitted values. 

The parameters describing the background proper time distributions are varied 
within their uncertainties. Background proper time distributions 
are parametrized using different methods, to check for 
possible systematic bias.
Different background samples are selected by varying the sideband regions, 
adding events from the lower sideband, which are excluded in 
the lifetime determination, or using events with wrong-sign correlations. 
Some cuts in the selection are varied to check the stability of the 
parametrizations. The shapes extracted from real data are compared with those
extracted from Monte Carlo events.
The resulting differences in the fitted lifetimes are used to estimate the  
systematic uncertainty.

For the remaining systematic errors, which are small compared to the statistical ones, the correlation between the lifetimes is not propagated into the ratio.

The systematic uncertainty due to the sample compositions is determined
by varying the branching fractions of Table~\ref{tab:b0br} within $\pm 1\sigma$
from the central values. In addition the uncertainty due to the assumption of isospin conservation has been estimated, allowing for a variation of 20\% relative to the exact symmetry.

\renewcommand{\arraystretch}{1.4}
\begin{table}[hbt]
\caption{\small Branching ratios used as input values in the
calculation of the sample composition.}
\begin{center}
\begin{tabular}{|l|c|c|} 
\hline
Decay & B.R. & Reference \\
\hline
$\bar{B}^0\to D^{\star+}\ell^-\bar{\nu}$ & 
$0.0460\pm0.0027$ & \cite{pdg98}\\
$\bar{B}^0\to D^{+}\ell^-\bar{\nu}$ &  
$0.0200\pm0.0025$ & \cite{pdg98} \\ 
$\bar{B}^0\to \ell^-\bar{\nu}X$ & 
$0.1045\pm0.0021$ & \cite{pdg98}\\
\hline
$
\hspace{-2mm}
\begin{array}{ll}
\bar{B}^0\to D^0\pi^+{\ell^-} \bar{\nu}\;+\;\\
\;\;\bar{B}^0\to D^{\star0}\pi^+{\ell^-} \bar{\nu}
\end{array}
$
& $0.0159\pm0.0036$ & \cite{ref:dss,ref:delphi_new_d} \\
\hline
${B}^-\to D^{\star+}\pi^-{\ell^-} \bar{\nu}$ &  
$0.0121\pm0.0018$ & \cite{ref:dss,ref:delphi_new_d} \\
\hline
${B}^-\to D^{+}\pi^-{\ell^-} \bar{\nu}$ &  
$0.0124\pm0.0048$ & \cite{ref:delphi_new_d} \\
\hline
$
\hspace{-2mm}
\begin{array}{ll}
{B}^-\to D^+\pi^-{\ell^-} \bar{\nu}\;+\;\\
\;\;{B}^-\to D^{\star+}\pi^-{\ell^-} \bar{\nu} 
\end{array}
$
& $0.0157\pm0.0031$ & \cite{ref:dss} \\
\hline
\end{tabular}
\end{center}
\label{tab:b0br}
\end{table}
%
%
%
%
%
%\begin{table}
%\caption{B branching fractions used in the lifetime measurement.}
%\label{tab:b0br}
%\begin{center}
%\begin{tabular}{|c|c|c|} \hline
%Decay & Branching ratio (\%) & Reference \\ \hline
%$ \bb \rightarrow \ds \ell^- \nu    $ & $ 4.50\pm0.45 $ &
%\protect\cite{ref:b1bl,ref:b1} \\
%$ \bb \rightarrow D^+ \ell^- \nu    $ & $ 1.9\pm0.5   $ & 
%\protect\cite{ref:b2} \\
%$ \bm \rightarrow \ds\pi^-\ell^-\nu $ & $ 0.96 \pm 0.33 $ & 
%\protect\cite{ref:dss} \\
%$ \bb \rightarrow \ell^- X            $ & $ 10.4\pm1.1 $  & 
%\protect\cite{ref:b1bl,ref:bl} \\  \hline
%\end{tabular}
%\end{center}
%\end{table}
%
%
% 

As explained in detail in \cite{ref:bl95}, a small fraction of four-body decays
where an extra pion is produced in the decay of the B meson contributes to
the sample composition, and is 
characterized by a different selection efficiency with respect to the real
signal.   The relative efficiencies
$$ \frac{\epsilon(B\ra\dords\pi\ell\nu)}{\epsilon(B\ra\dords\ell\nu)}=0.75\pm 0.10 ~(\mbox{\rm for $\bar B^0$}) ~~~~
0.64\pm 0.10 ~(\mbox{\rm for $B^-$})$$
therefore enter into the calculation and this uncertainty (coming from the limited Monte Carlo statistics) is propagated to the measured lifetimes. 
 
The parameters of the decay length resolution function 
are varied within their
errors to estimate the resulting uncertainty \cite{ref:blgls}.

\begin{table}
\caption{\small Sources of systematic error on the fitted lifetimes.}
\label{tab:syst}
\begin{center}
\begin{tabular}{|l|c|c|c|} \hline
Source of error &
\multicolumn{3}{c|}{Contribution to systematic error} \\ \cline{2-4}
  & \t0 \ (ps) & \tp\ (ps) & $\tp /\t0 $ \\  \hline
B momentum reconstruction  &
  $\pm 0.025$ & $\pm 0.026 $ & $\pm0.009$  \\ \hline
Background treatment&  
  $\pm0.020$ & $\pm0.020$ & $\pm0.010$  \\ \hline
Sample compositions &  $\pm0.003$ & $\pm0.003$ & $\pm0.004$ \\ \hline
\dpil\ relative efficiency & 
  $\pm0.006$ & $\pm0.006$ & $\pm0.006$  \\ \hline
Decay length resolution &  
              $\pm0.008$ & $\pm0.008$ & $\pm0.008$ \\ \hline \hline
Total  & $\pm 0.034$& $\pm 0.035$ & $\pm0.018$  \\ \hline
\end{tabular}
\end{center}
\end{table}
%
%\end{document}
%\input calvet
%
%\input oest
%
\section{Conclusions}
The lifetimes of the charged and neutral $B$ mesons have been measured
with the full statistics collected by the \A\ detector at and around the
Z peak energy. The data sample was recently reprocessed,
achieving improved tracking performance.
A maximum likelihood fit to the proper time distributions 
of 1880 \dsl\ and 2856 \dl\ candidates 
yields the following results for the \b0 \ and \bp\ lifetimes and
their ratio:
\begin{eqnarray*}
\t0 & =  & 1.518 \pm 0.053\pm 0.034\mathrm{\ ps,} \\
\tp & =  & 1.648 \pm 0.049\pm 0.035 \mathrm{\ ps,} \\
\tp/\t0 &= & 1.085 \pm 0.059 \pm 0.018 ~,
\end{eqnarray*}
where the first error is statistical and the second is systematic. 

\noindent
These results supersede the corresponding ones of \cite{ref:bl95}.
Averaging with the results of the other methods presented in \cite{ref:bl95} the combined values are:
\begin{eqnarray*}
\t0 & =  & 1.496 \pm 0.048 \pm 0.033\mathrm{\ ps,} \\
\tp & =  & 1.644 \pm 0.048 \pm 0.034\mathrm{\ ps,} \\
\tp/\t0 &= & 1.104 \pm 0.057 \pm 0.019.  
\end{eqnarray*}

\section*{Acknowledgments}
We thank our colleagues in the accelerator divisions for the excellent 
performance of LEP. Thanks also to the many engineering and 
technical personnel at CERN and at the home institutes for their contributions
to the performance of the \A\ detector. Those of us from non-member 
states thank CERN for its hospitality.
 
%\nopagebreak
%\clearpage

\end{document}